# Superconducting Micro- and Nanohelices

Vladimir M. Fomin*,[†], Roman O. Rezaev[‡,§], Evgenii A. Levchenko[§], Daniel Grimm[†,#], and Oliver G. Schmidt[†,#]

[†]*Institute for Integrative Nanosciences, IFW-Dresden, Helmholtzstraße 20, D-01069 Dresden, Germany*

[‡]*Moscow Engineering Physics Institute, Kashirskoye Shosse 31, Moscow, 115409, Russia*

[§]*Tomsk Polytechnic University, Lenin ave. 30, Tomsk, 634050, Russia*

[#]*Material Systems for Nanoelectronics, Technische Universität Chemnitz, Reichenhainer Straße 70, D-09107 Chemnitz, Germany*

ABSTRACT: Superconducting micro- and nanohelices are proposed for the first time. A theoretical investigation of the superconducting state in the helical coils at the micro- and nanoscale is performed within the time-dependent Ginzburg-Landau approach. The pattern and number of vortices in a stationary distribution are determined by their confinement to the ultrathin helical spiral and can therefore be efficiently controlled by the helical stripe width and the helical pitch distance for both dense and sparse helices. Quasi-degeneracy of vortex patterns is manifested in the helical spiral when the number of vortices is incommensurable with the total number of half-turns. With increasing radius, superconducting helical spirals provide a physical realization of a transition from the vortex pattern peculiar to an open tube to that of a planar stripe.

1. Introduction

Due to the *chiral geometry*, helical nanoarchitectures provide a significant advancement in modern nanosciences and nanotechnologies. Helical structures have been synthesized of various materials, including carbon-based [1 to 3], all-semiconductor [4], metal [5, 6], DNA-based chiral plasmonic nanostructures [7], hybrid metal-semiconductor [8] nanocoils and nanocoils of semiconductor oxides [9]. Because of their unique morphology as well as mechanical, electrical, magnetic and optical properties [10], nanohelices have attracted interest for application as handedness-switchable chiral metamaterials [5, 11], electromagnetic actuators and sensors [12], micro- and nano-electromechanical systems [13], nanorobotics [14], sperm-carrying micromotors [15], nanoscale elastic energy storage [16], photodetectors [17], smart electric conductors, magnetic sensors and electromagnetic wave absorbers [18].

Fabrication, characterization and application of spiral-shaped structures are of central interest for the advancement of *superconductor* technologies. The main attention in this field has been attracted to planar (2D) structures. For instance, X-ray [19] and single-photon [20, 21] detectors based on superconducting nanowire planar spirals have developed into a mature technology [22]. Spiral antennas are shown to produce a very effective coupling between hot electron bolometers [23]. Superconducting spiral resonators have revealed potential for constructing one-, two-, and three-dimensional metamaterials [24]. Superconducting metamaterials made up of planar spiral Nb elements have provided minimized Ohmic losses, compact design with high quality factor and sensitive tuning of resonant frequency via temperature and magnetic field [25]. The rf current distributions in Nb planar spiral resonators along the width of the individual turns of the resonators reveals an unconventional behavior: the maximum current is observed in the middle of the structure, which is associated with the geometry and the cancellation of magnetic field between the turns [26]. This is favorable for handling high powers and thus expands the range of applicability for rf/microwave resonators.

*Macroscopic* superconducting tapes of various materials (from $Nb_3Sn$ [27] to high-$T_c$ [28-30]) helically wound around a cylinder have been intensively explored with the aim of reducing losses and cost of power transmission cables. The study of self-field hysteresis losses of helicoidal superconductor structures has demonstrated that the shape of the twisted wire has influence on the qualitative loss behavior [31]. Investigations of 3D helical spiral superconducting *micro-* and *nanostructures* remain a challenge for both experiment and theory. Superconducting micro- and nanohelices are considered in the present paper for the first time. The heuristic value of such studies is implied by the fact that nano- and microstructuring is one of the main avenues of the modern advancement of superconductor physics and technology, including superconducting electronics [32], radiation modulators in the THz or sub-THz range [33], and superconducting qubits for quantum computing [34].

Strain-driven self-assembly of rolled-up architectures on the nano- and microscale [4, 35] allows for fabrication of spiral-shaped Swiss roll micro- and nanotubes with superconductor layers (e.g., InGaAs/GaAs/Nb) [36, 37]. Those hybrid structures open hitherto unprecedented possibilities for experimental investigation of vortex matter in superconductors with curved geometries. In the rolled-up structures, the quasi-two-dimensionality of the superconducting layer is combined with a curvature. Vortex equilibrium distributions as well as vortex dynamics are significantly influenced by the curvature of the superconductor at the nano- or microscale [38, 39]. The roll-up technique allows, further, for *combining curvature at the nanoscale with chirality* by fabricating micro- and nanocoils [40, 41]. The present paper is aimed at a quantitative analysis of the synergetic effects of curvature and chirality on vortex equilibrium distributions on superconductor helical nanocoils. It is organized as follows. A physical model of a superconducting helical spiral coil is presented in Sec. 2. Calculation of stationary superconducting vortex patterns in helical spiral coils using the finite-difference scheme for the time-dependent Ginzburg-Landau equation is described in Sec. 3. Analysis of the vortex patterns occurring for different magnetic fields and geometrical parameters of a helical spiral coil in Sec. 4 is followed by Conclusions.

## 2. Model

We consider a helical coil of radius $R$, width $W$ and pitch distance $P$ with $N=2$ windings shown in Fig. 1(a). The helix has the total length $L = N\sqrt{(2\pi R)^2 + P^2}$ and the helix angle $\theta = \tan^{-1}(2\pi R/P)$. A detailed geometrical model of the helical spiral coil is represented in Appendix A.

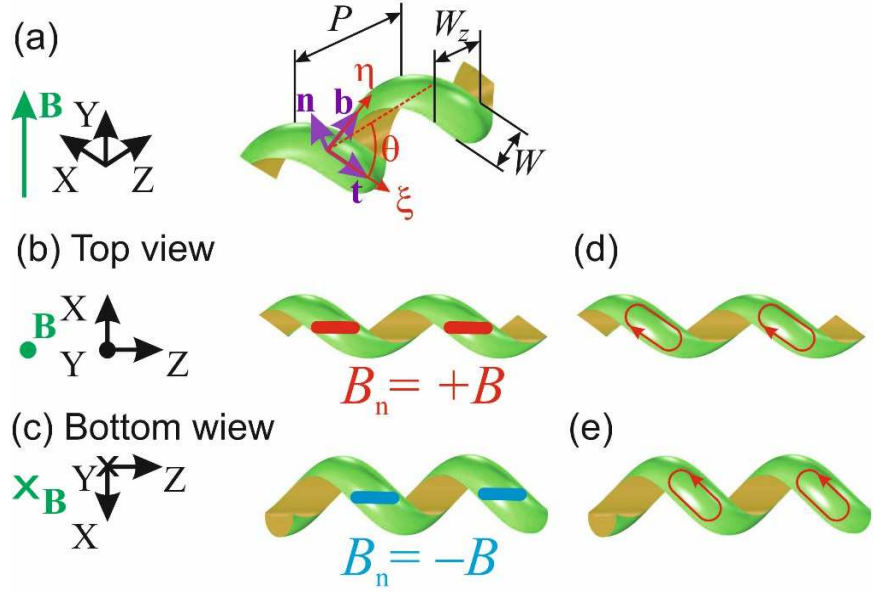

**Figure 1**. (a) Schematics and geometrical parameters (see the text) of a helical spiral coil. The spiral axis is along the z-axis. The magnetic field **B** is along the y-axis. At any point of the helical surface **t**, **n**, **b** denote, correspondingly, tangential, normal and binormal unit vectors. (b) [(c)] Scheme of the areas with the maximal [minimal] values of the normal component of the magnetic field $B_n=B$ [$B_n=-B$]. The linear extent of those areas along the z-axis is of the order of $W_z$. (d) [(e)] Scheme of a set of screening currents in the top [bottom] parts of the spiral turns.

Under the assumption of a thin helical coil, the superconducting order parameter $\psi$ is governed by the Ginzburg-Landau equation, which will be used in the following dimensionless form [42]:

$$\frac{\partial \psi}{\partial t} = (\nabla - i\mathbf{A})^2 \psi + 2\kappa^2 \psi \left(1 - |\psi|^2\right), \qquad (1)$$

where κ is the Ginzburg-Landau parameter and **A**=−$Bx$**e**$_z$ is the vector-potential of the uniform magnetic field **B**=$B$**e**$_y$ (see Fig. 1a). It is complemented with the boundary conditions on the free boundaries

$$(\nabla - i\mathbf{A})\psi\big|_{n,\text{boundary}} = 0, \quad (2)$$

where "*n*" denotes a component along the normal to the boundary of the helical stripe. The external normal to the boundary of the helical stripe coincides with [is opposite to] the binormal **b** on the boundary (ξ, $W$/2) [(ξ, −$W$/2)], for instance,

$$(\nabla - i\mathbf{A})\psi \cdot \mathbf{b}(\xi, \pm W/2)\big|_{\text{boundary}} = 0 \text{ at any } \xi. \quad (3)$$

It coincides with [is opposite to] the tangential vector **t** on the boundary ($L$,η) [(0,η)], for instance,

$$(\nabla - i\mathbf{A})\psi \cdot \mathbf{t}(L \text{ or } 0, \eta)\big|_{\text{boundary}} = 0 \text{ at any } \eta. \quad (4)$$

An infinitesimally thin *helical stripe* of width $W$ is a two-parametric surface represented by the radius-vector (see Appendix B)

$$\mathbf{r}(\xi,\eta) = \left\{ R\cos\left(\frac{2\pi}{\ell}s\right), R\sin\left(\frac{2\pi}{\ell}s\right), \xi\cos\theta + \eta\sin\theta \right\}, \quad s = \xi - \eta\cot\theta, \quad (5)$$

Here the coordinates $\xi \in [0,L]$ and $\eta \in [-W/2, W/2]$ are counted along and across the helical stripe, correspondingly. The natural orthogonal coordinates $(\xi, \eta)$ form a convenient basis for solving the Ginzburg-Landau equation on a helical spiral (see Appendix C).

The stationary vortex distributions in a helical spiral emerge as a result of the strongly inhomogeneous normal magnetic field, whose pattern, shown in Fig. 2(a) in the coordinates $(\xi, \eta)$ depends on the geometrical characteristics of the spiral stripe. The normal component is maximal ($B_n = B$) in the areas, where the unit normal vector $\mathbf{n}(\xi, \eta)$ is parallel or close to the direction of the applied magnetic field $\mathbf{e}_y$ [red in Fig. 1(b)], and minimal ($B_n = -B$) in the areas, where the normal vector is antiparallel or close to the direction opposite to the applied magnetic field [blue in Fig. 1(c)]. The above areas are favourable for the occurrence of vortices (compare with Ref. [38]).

## 3. Calculation of Stationary Vortex Patterns

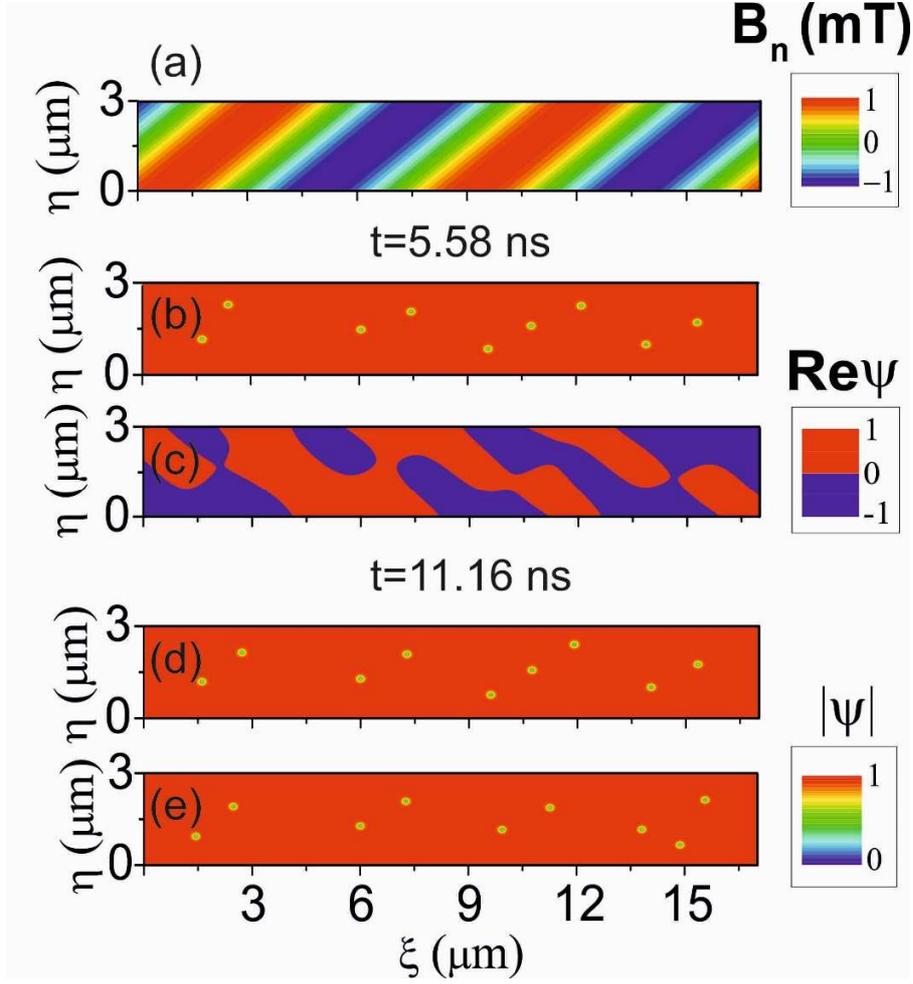

**Figure 2**. (a) Distribution of the normal component of the magnetic field $B_n = \mathbf{B} \cdot \mathbf{n}(\xi,\eta)$ in a helical coil with parameters: $R$=0.96 μm, $W$=3 μm, $P$=6 μm, $L$=2$\ell$=17.02 μm. The applied uniform magnetic field is $B = 1$ mT. (b,d,e) Distribution of the modulus of the order parameter $|\psi|$ in the helical coil calculated using the orthogonal mesh [see Eq. (B4) in Appendix B] with $N_\xi$= 481 and $N_\eta$= 241. (c) Distribution of the real part of the order parameter Re$\psi$. Data in (b) and (c) are obtained at $t$ = 5.58 ns, while those in (d) and (e) at $t$ = 11.16 ns. A stochastic potential, which enables transitions of the vortex system between patterns with close or equal energies (see Appendix D), has a higher magnitude in (e), than in (d).

Calculation of the order parameter is performed using the finite-difference scheme (see [42], where it was developed for planar nanostructured superconductors) for the time-dependent Ginzburg-Landau equation (1) with the boundary conditions (4), (5) realized on the orthogonal mesh [see Eq. (B4) in Appendix B]. The materials parameters for nanostructured Nb at the relative temperature $T/T_c=0.95$ are as follows: the coherence length $\xi=56$ nm, the penetration depth $\lambda=279$ nm, the Ginzburg-Landau parameter $\kappa=5$, the diffusion coefficient $D=11.2\times10^{-4}$ m$^2$s$^{-1}$ [38]. The initial distribution of the complex order parameter is taken as a random complex field. An ultrasmall stochastic potential is added to the shifted Laplace operator in Eq. (1), in order to allow for transitions of the vortex system between different configurations with close or equal energies and to facilitate the evolution of the order parameter to a state with a minimal free energy (see Appendix D). The evolution of the order parameter is traced towards a stationary state. As illustrated in Figs. 2(b) and 2(d), the stationary distribution of the order parameter contains single vortices or chains of vortices aligned along the lines of the maximal or minimal normal component of the magnetic field [compare Figs. 2(b) and 2(d) with Fig. 2(a)]. The fact that the lines of the $2\pi$-phase shift of the order parameter [Fig. 2(c)] go through dips of the modulus of the order parameter [Figs. 2(b)] evidences that those dips represent *superconducting vortices*.

Given the magnetic field, the pattern and number of vortices are determined by their confinement to the helical spiral. The boundary conditions (4) are specific for the helical spiral and provide a major distinction from the case of an open cylinder [38]. Qualitatively, a certain distribution of vortices on the surface of a cylinder is established due to the vortex-vortex interaction under the condition that screening currents flow close to the boundaries of the upper and the lower halves of the cylinder with respect to the direction of the applied magnetic field. In open superconductor tubes, screening currents flow in each half of *the whole cylinder* as shown in Fig. 1(b) of [38]. As distinct from that, in the helical coil those currents flow in each *half-turn* [as shown schematically in Figs. 1(d) and 1(e)], and therefore the confinement of vortices to the helical stripe affects the vortex pattern formation through a *longer boundary* than that in a cylinder of the same radius and comparable overall height.

The different numbers of vortices occurring in different half-turns (for example, two and three at $B = 1$ mT) is attributed to the adjustment of the vortex matter to such a magnetic field, which requires a number of vortices, which is *incommensurable* with (in other words, is not an integer multiple of) the total number of half-turns. In the course of evolution, alternative patterns of vortices distributed over half-turns may occur. At $B = 1$ mT, there are different configurations of nine vortices with three vortices in different half-turns, shown, for instance, in Figs. 2(d) and 2(e). This is a manifestation of a *quasi-degeneracy* of vortex patterns in a helical spiral. Henceforth, we describe the vortex distributions by the *average* number of vortices per half-turn.

### 4. Discussion of Vortex Patterns

A quantitative analysis of the vortex patterns in helical coils for different magnetic fields and geometrical parameters of a helical coil is represented in what follows.

When increasing the magnetic field, the pattern of the vortex distribution changes from single linear chains aligned along the lines of the maximal or minimal values of $B_n$ [see the case $B = 1$ mT in Figs. 2(b) and 2(d)] towards a few chains represented in Fig. 3, (see the cases $B = 5$ and 10 mT). These vortex chains are concentrated in the areas of the maximal or minimal values of $B_n$, as sketched in Figs. 1(b) and 1(c). This pattern of ordering emerges as a result of the interplay between the vortex confinement to the above-mentioned areas in helical stripes and the vortex-vortex interaction.

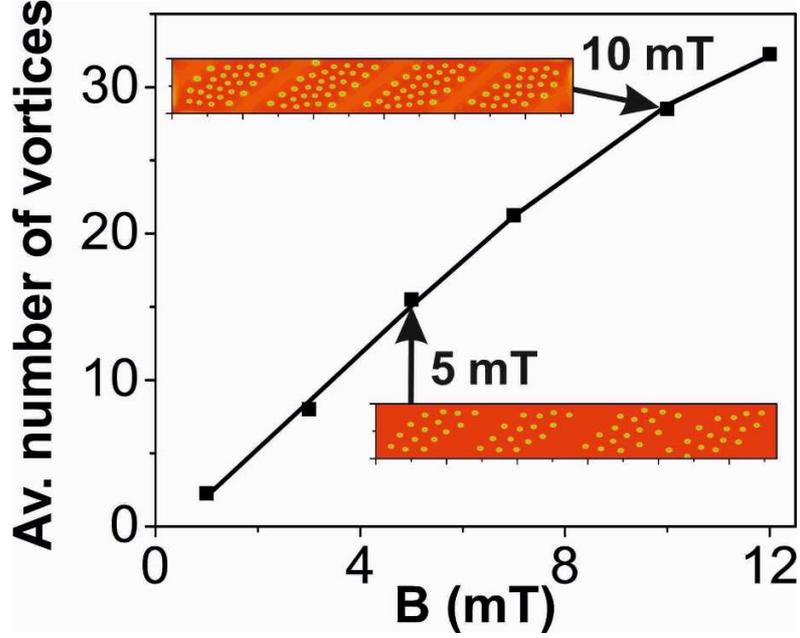

**Figure 3**. Average number of vortices per half-turn as a function of the applied uniform magnetic field for a helical coil with the same geometrical parameters as those in Fig. 2. In insets, distributions of the order parameter are shown for $B = 5$ mT and 10 mT. The color scale is the same as in Fig. 2(b). Data are obtained at $t = 5.58$ ns. (In Figs. 3 to 7, calculated data are represented with squares, while solid lines are guides to the eye.)

We start the systematic study of the impact of geometry on the distribution of vortices with a sparse helical spiral, when the pitch distance appreciably exceeds the value $P_{overlap}$, so that there is a significant distance between neighbouring turns. The average number of vortices per half-turn increases with $W$, as illustrated in Fig. 4. At a low magnetic field $B = 1$ mT the average number of vortices per half-turn with increasing helical stripe width tends to a slow function of $W$ due to the developing screening currents. With rising magnetic field up to $B = 5$ mT, the increase becomes faster due to developing multiple-chain patterns. At even higher magnetic fields ($B > 5$ mT), the dependence approaches a linear function, which reflects the geometrical origin of the effect: in a sparse helical spiral with a fixed pitch distance, the magnetic flux through the helical stripe in a fixed magnetic field rises directly proportional to its width. The vortices occur even in sparse helical spirals with rather narrow stripes ($W \sim 1$ μm), because the effective width of a stripe along the spiral axis $W_z$ [see Eq. (B2) in Appendix B] increases with $P$.

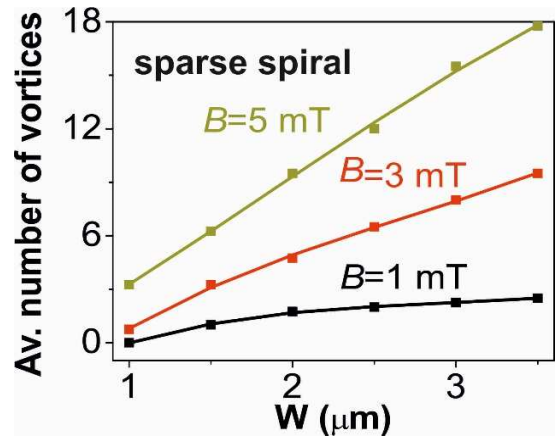

**Figure 4**. Average number of vortices per half-turn as a function of the helical stripe width for a sparse spiral at $P = 6$ μm placed in a fixed magnetic field ($B = 1$ mT, 3 mT and 5 mT). Data are obtained at $t = 5.58$ ns.

We consider further a dense helical spiral, when the pitch distance $P$ slightly exceeds the value $P_{overlap}$. [seeEq. (B3) in Appendix B], at which the subsequent turns start to overlap. In this case, the neighbouring turns are close to (almost touching) each other. The distribution of vortices occurs to be very sensitive to the helical stripe width $W$. In Fig. 5, the average numbers of vortices per half-turn in dense helical spirals are represented as a function of the helical stripe width $W$. The increase of the average number of vortices per half-turn with increasing width $W$ occurs faster than the linear dimension (of the order of $W_z \approx W$) of the areas with the maximal/minimal values of $B_n$. This nonlinear behaviour can be again attributed to developing multiple-chain patterns, as shown in the insets to Fig. 3.

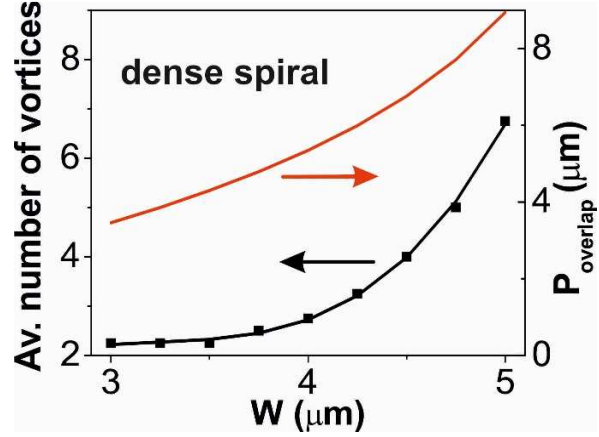

**Figure 5**. Average number of vortices per half-turn as a function of the helical stripe width for a dense spiral. At any value of $W$, the pitch distance slightly exceeds the value $P_{overlap}$ (also plotted). Data are obtained at $t = 5.58$ ns. The applied uniform magnetic field is $B = 1$ mT.

The increase of the average number of vortices with the pitch distance depends on the applied magnetic field. For single-chain patterns at weaker fields, as shown in Fig. 6 for $B = 1$ mT, the increase is slow presumably because of an appreciable impeding of the vortex nucleation due to the boundary conditions at the helical stripe edges. For multiple-chain patterns at stronger fields, the increase of the average number of vortices per half-turn occurs faster than the linear dimension ($\sim W_z$) of the areas with the maximal/minimal values of $B_n$. This behaviour may be attributed to the following *combination of geometrical and physical factors*: with increasing $W_z$, not only the magnetic flux penetrating the helical spiral increases, but impeding of the vortex nucleation due to the boundary conditions at the helical stripe edges becomes relatively weaker. A stepwise increase of the average number of vortices as a function of the pitch distance in Fig. 6 reflects the geometric fact that in the helical coil with 2 turns, a nucleation or denucleation of a vortex changes the average number of vortices per half-turn by 0.25 or – 0.25. In helical coils with a larger number $N$ of turns, the magnitude of the smallestchangeof the average number of vortices decreases as $1/(2N)$.

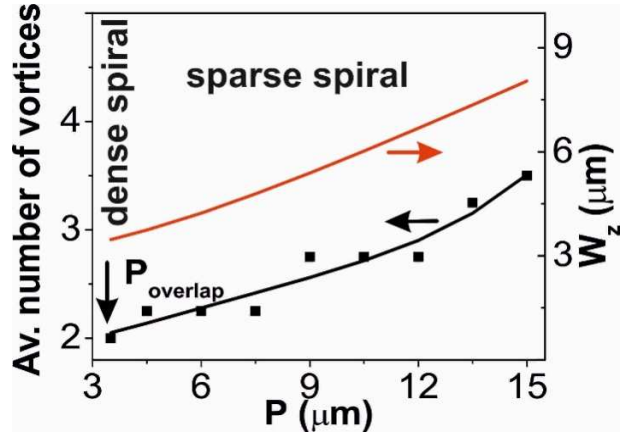

**Figure 6**. Average number of vortices per half-turn as a function of the pitch distance at the helical stripe width $W = 3$ μm. The effective width of a stripe along the spiral axis $W_z$ is also plotted. The applied uniform magnetic field is $B = 1$ mT. Data are obtained at $t = 16.74$ ns.

The average number of vortices per half-turn increases with the radius, as shown in Fig. 7. With rising magnetic fields from $B = 1$ mT to $B = 5$ mT, the increase becomes faster due to developing of larger multiple-chain fragments.This occurs because of the following geometric factor. When increasing the radius $R$at fixed values of the helical stripe width $W = 3$ μmand the pitch distance$P = 6$ μm, the areas with maximal [minimal] value of the normal component of the magnetic field $B_n=B$ [$B_n=-B$](marked with red in Fig. 1(d) [blue in Fig. 1(e)]) expand proportionally to the increasing length of the stripe along the centreline $L$.The insets to Fig. 7 represent vortex patters at $B= 5$ mT for radii $R = 0.72$ μm,1.68 μmand 2.4 μm. For a smaller radius ($R = 0.72$ μm), vortex chains are arranged parallel to the lines wherethe normal component of the magnetic field has its maximal [minimal] value of $B_n=B$ [$B_n=-B$]. With increasing radius, the average number of vortices per half-turn rises more prominently at higher magnetic field values, similarly to Fig. 3. At $B = 1$ mT and 3 mT this behaviour is manifested over the whole interval of the considered radii. However, at still higher magnetic fields, e.g., $B = 5$ mT, the average number of vortices per half-turn as a function of radius saturates. This trend of vortex pattern in each of the upper and lower areas of the helical stripe reflects behaviour similar to that in a planar stripe perpendicular to an applied magnetic field, corresponding to the limit of negligible curvature. The multi-vortex pattern acquires the shape of vortex chains parallel to the sides of the helical stripe (inset for $R = 2.4$ μm). This may be attributed to stronger screening currents, which effectively repel vortices from the sides of the stripe in helical spirals of a larger radius. At intermediate radii, an interplay of both trends of vortex ordering is observed: (i) parallel to the lines where the normal component of the magnetic field has its maximal [minimal] value, which is typical of an open tube [38], and (ii) parallel to the sides of the helical stripe, specific for a planar stripe [43]. The interplay leads to a mixed pattern even with some signs of a hexagonal order (inset for $R = 1.68$ μm).

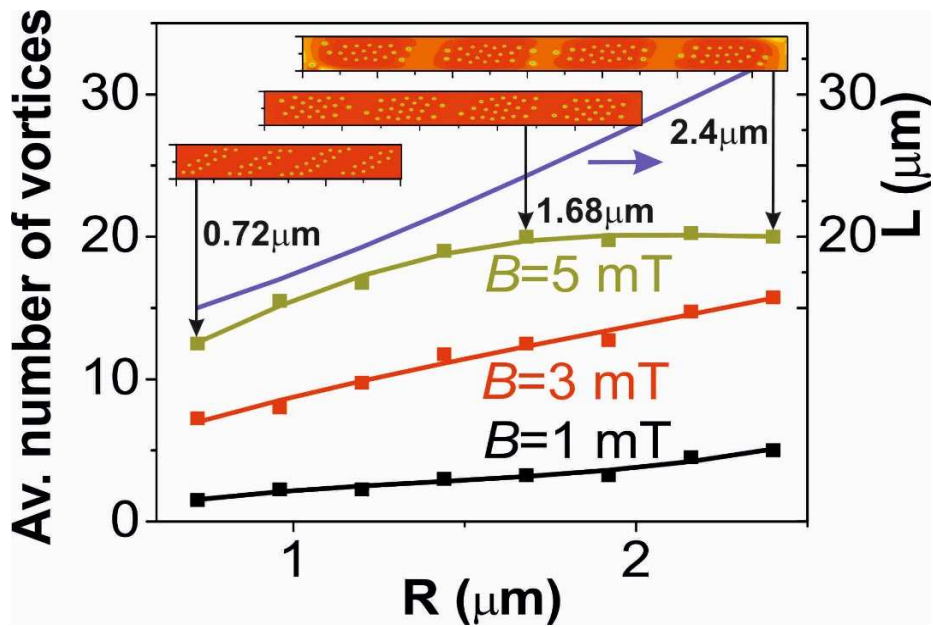

**Figure 7**. Average number of vortices per half-turn as a function of the radius at the helical stripe width $W = 3$ μm and the pitch distance $P = 6$ μm for different values of the applied uniform magnetic field ($B = 1$ mT, 3 mT and 5 mT). Data are obtained at $t = 16.74$ ns. The length of the stripe along the centerline $L$ is also plotted. Distributions of the order parameter are shown in the insets for $R = 0.72$ μm, 1.68 μm and 2.4 μm.

## 5. Conclusion

The pattern and number of vortices in a stationary distribution are determined by their confinement to the helical spiral. For a dense helical spiral, as well as for a sparse one at lower magnetic fields, the increase of the average number of vortices with increasing the stripe width occurs faster than the growth in the linear dimension of the areas with extremal values of the normal component of the magnetic field. This behaviour can be presumably attributed to the trend of changing the ordering patterns from one to a few chains of vortices. For a sparse helical spiral at higher magnetic fields, the average number of vortices appears to be a linear function of the stripe width, what reflects the geometrical origin of the effect. The increase of the average vortex number with increasing pitch distance occurs faster than the growth in the linear dimension of the areas, where the normal component of the magnetic field has its extremal values, due to an interplay of geometrical and physical factors. The revealed quasi-degeneracy of the vortex patterns generates very interesting dynamics of vortices in helical spirals. Superconducting helical coils of increasing radius provide a physical realization of a transition from the vortex pattern of an open tube to that of a planar stripe. The revealed excellent tunability of the superconducting vortex patterns makes micro- and nanohelices highly promising for application in superconducting electronics, radiation modulators and superconducting qubits.

**Acknowledgement**

We gratefully acknowledge discussions with R. Wördenweber, R. R. Dusaev, D. Bürger, S. Lösch and V. Engemaier. This work has been supported by the bilateral BMBF-Russia research grant 01DJ13009.

## APPENDIX A: Geometrical model

For the helical coil shown in Fig. 1(a), the centerline in the Cartesian coordinates X, Y, Z is

$$\mathbf{r}(\xi) = \left\{ R\cos\left(\frac{2\pi}{\ell}\xi\right), R\sin\left(\frac{2\pi}{\ell}\xi\right), \frac{P}{\ell}\xi \right\}, \tag{A1}$$

where $R$ is the radius of the cylinder, on which the coil is wound up, $P$ is the helix pitch distance,

$$\ell = \sqrt{(2\pi R)^2 + P^2} \tag{A2}$$

is the length of a single turn, the (longitudinal) coordinate along the helix $\xi \in [0, L]$, $L$ is the total length $L$ of the centerline. The helix angle $\theta$, which is defined through the relations

$$\frac{2\pi R}{\ell} = \sin\theta, \quad \frac{P}{\ell} = \cos\theta, \quad \Rightarrow \quad \theta = \tan^{-1}\frac{2\pi R}{P}, \tag{A3}$$

allows for the following representation of the centerline (A1):

$$\mathbf{r} = \left\{ \frac{\ell}{2\pi} \sin\theta \cos\left(\frac{2\pi}{\ell}\xi\right), \frac{\ell}{2\pi} \sin\theta \sin\left(\frac{2\pi}{\ell}\xi\right), \xi\cos\theta \right\}. \tag{A4}$$

## APPENDIX B: Natural orthogonal coordinates

An infinitesimally thin *helical stripe* of width $W$ is a two-parametric surface represented by the radius-vector

$$\mathbf{r}(\xi,\eta) = \left\{ R\cos\left(\frac{2\pi}{\ell}s\right), R\sin\left(\frac{2\pi}{\ell}s\right), \xi\cos\theta + \eta\sin\theta \right\}, \quad s = \xi - \eta\cot\theta, \tag{B1}$$

where the (transverse) coordinate across the helix $\eta \in [-W/2, W/2]$. The width of the helical stripe measured along the z-axis [see Fig. 1(a) of the main text] is

$$W_z = W/\sin\theta = W\ell/(2\pi R). \tag{B2}$$

The condition of no overlap of the subsequent turns $P > W_z$ is represented in the form

$$P > P_{overlap} \text{ with } P_{overlap} = \left(\frac{1}{W^2} - \frac{1}{(2\pi R)^2}\right)^{-1/2}. \tag{B3}$$

The $N_\xi \times N_\eta$ mesh for the finite-difference calculation is naturally parameterized as follows:

$$\mathbf{r}_{ij} = \mathbf{r}(\xi_i, \eta_j) =$$
$$\left\{ R\cos\left(\frac{2\pi}{\ell} s_{ij}\right), R\sin\left(\frac{2\pi}{\ell} s_{ij}\right), \xi_i \cos\theta + \eta_j \sin\theta \right\},$$
$$s_{ij} = \xi_i - \eta_j \cot\theta, \qquad (B4)$$
$$\xi_i = \frac{i-1}{N_\xi - 1} L, \quad i = 1,\ldots,N_\xi;$$
$$\eta_j = \frac{j-1}{N_\eta - 1} W - \frac{W}{2}, \quad j = 1,\ldots,N_\eta.$$

The unit tangential vector along the centerline is

$$\mathbf{t}(\xi,\eta) = \frac{\partial \mathbf{r}(\xi,\eta)}{\partial \xi} = \left\{ -\sin\theta \sin\left(\frac{2\pi}{\ell} s\right), \sin\theta \cos\left(\frac{2\pi}{\ell} s\right), \cos\theta \right\}, \quad (B5)$$

the unit normal vector to the cylindrical surface is

$$\mathbf{n}(\xi,\eta) = \frac{\partial \mathbf{r}(\xi,\eta)}{\partial R} = \left\{ \cos\left(\frac{2\pi}{\ell} s\right), \sin\left(\frac{2\pi}{\ell} s\right), 0 \right\}, \quad (B6)$$

and the binormal unit vector $\mathbf{b}(\xi,\eta) = \mathbf{n}(\xi,\eta) \times \mathbf{t}(\xi,\eta)$ is

$$\mathbf{b}(\xi,\eta) = \left\{\cos\theta\sin\left(\frac{2\pi}{\ell}s\right), -\cos\theta\cos\left(\frac{2\pi}{\ell}s\right), \sin\theta\right\}. \quad (B7)$$

The unit vectors **t**, **n** and **b** are shown in Fig. 1(a).

## APPENDIX C: Boundary conditions for the Ginzburg-Landau equation on a helical stripe

In terms of the cylindrical coordinates

$$\varphi = \frac{2\pi s}{\ell}, \quad z = \xi\cos\theta + \eta\sin\theta \quad (C1)$$

the helical stripe is represented as a part of the cylindrical surface:

$$\mathbf{r}(\varphi, z) = \{R\cos(\varphi), R\sin(\varphi), z\}. \quad (C2)$$

In the cylindrical coordinates the shifted Laplace operator has the form

$$(\nabla - i\mathbf{A})^2 = \frac{1}{R^2}\frac{\partial^2}{\partial\varphi^2} + \left(\frac{\partial}{\partial z} - iA\right)^2. \quad (C3)$$

Substituting the definition $s = \xi - \eta\cot\theta$ into (C1), we arrive at the *transformation of coordinates*

$$R\varphi = (\xi\sin\theta - \eta\cos\theta), \quad z = \xi\cos\theta + \eta\sin\theta. \quad (C4)$$

An *inverse transformation* can be interpreted as a rotation of the cylindrical coordinates by the helix angle:

$$\xi = R\varphi\sin\theta + z\cos\theta, \quad \eta = -R\varphi\cos\theta + z\sin\theta. \quad (C5)$$

The coordinate axes $(\xi, \eta)$ are represented in Fig. 1(a). Under the rotation (C5), the shifted Laplace operator (C3) transforms as follows:

$$(\nabla - i\mathbf{A})^2 = \frac{\partial^2}{\partial \xi^2} + \frac{\partial^2}{\partial \eta^2} - 2iA\left(\frac{\partial}{\partial \xi}\cos\theta + \frac{\partial}{\partial \eta}\sin\theta\right) - A^2 \quad (C6)$$

is represented in terms of the longitudinal and transverse components of the vector-potential

$$A\cos\theta = A_\xi \quad \text{and} \quad A\sin\theta = A_\eta, \quad (C7)$$

correspondingly:

$$(\nabla - i\mathbf{A})^2 = \left(\frac{\partial}{\partial \xi} - iA_\xi\right)^2 + \left(\frac{\partial}{\partial \eta} - iA_\eta\right)^2. \quad (C8)$$

In terms of the longitudinal and transverse components of the vector-potential (C7) the boundary conditions on two pairs of boundaries are ***explicitly independent*** from each other:

$$\left(\frac{\partial}{\partial \eta} - iA_\eta\right)\psi(\xi, \pm W/2)\bigg|_{\text{boundary}} = 0 \quad \text{at any } \xi; \quad (C9)$$

$$\left(\frac{\partial}{\partial \xi} - iA_\xi\right)\psi(L \text{ or } 0, \eta)\bigg|_{\text{boundary}} = 0 \quad \text{at any } \eta. \quad (C10)$$

This property of the boundary conditions makes the coordinates introduced by the equation (C5) particularly useful for solving the Ginzburg-Landau equation on a helical spiral.

## APPENDIX D: Stochastic potential

The Ginzburg-Landau equation of Eq. (1)

$$\frac{\partial \psi}{\partial t} = \mathbf{L}(\psi), \quad \mathbf{L}(\psi) = (\nabla - i\mathbf{A})^2 \psi + 2\kappa^2 \psi\left(1 - |\psi|^2\right) \quad (D1)$$

with a differential operator $\mathbf{L}(\psi)$ is solved using a finite-difference space-time scheme. The evolution of the order parameter during one step in discrete time, from $t_n$ to $t_{n+1}$, is obtained in the form

$$\psi(\mathbf{r}_{ij}, t_{n+1}) \approx \psi(\mathbf{r}_{ij}, t_n) + \mathbf{L}_{ij}[\psi(\mathbf{r}_{kl}, t_n)](t_{n+1} - t_n) + A\left\{[\text{rand}_1(\mathbf{r}_{ij}) - 0.5] + i[\text{rand}_2(\mathbf{r}_{ij}) - 0.5]\right\}, \quad (D2)$$

where $\mathbf{L}_{ij}[\psi(\mathbf{r}_{kl},t_n)]$ is a matrix functional generated by the differential operator $\mathbf{L}(\psi)$ on the mesh (B4). The third term in the right-hand side contains an amplitude $A$ and two mutually independent random functions: $\text{rand}_1(\mathbf{r})\in[0,1]$ and $\text{rand}_2(\mathbf{r})\in[0,1]$. It represents a stochastic potential, which is added to the operator $\mathbf{L}(\psi)$ in order to allow for transitions of the vortex system between different configurations with close or equal energies. This stochastic potential prevents a "freezing" of the order parameter configuration in a metastable state by pushing it out of any state with a local free-energy minimum achieved in the course of the numerical solution. Therefore the stochastic potential facilitates the evolution of the order parameter to a stable state with a minimal free energy. Calculations performed with two different values of the amplitude, $A=10^{-5}$ and $A=10^{-4}$, are shown in Figs. 2(d) and (e), correspondingly.